\documentclass[11pt]{article}

\usepackage[letterpaper,top=3cm,bottom=3cm,left=3cm,right=3cm,marginparwidth=1.75cm]{geometry}

\usepackage[english]{babel}
\usepackage{amsmath}
\usepackage{graphicx}
\usepackage[colorlinks=false, allcolors=blue]{hyperref}
\usepackage[section]{placeins}
\usepackage{booktabs}  
\usepackage{caption}
\usepackage{apacite}
\usepackage{natbib}
\usepackage{setspace}

\begin{document}

\title{\vspace{-50pt}Primate-like perceptual decision making emerges through \\ deep recurrent reinforcement learning}
\date{}
\maketitle

\vspace{-60pt}

\noindent
\begin{center}
\author{Nathan~J. Wispinski \textsuperscript{1,2,*}, Scott~A. Stone \textsuperscript{1,2}, Anthony Singhal \textsuperscript{1,2}, \\
Patrick~M. Pilarski \textsuperscript{3,4}, \& Craig~S. Chapman \textsuperscript{2,5}}
 \\

\vspace{10pt}

1. Department of Psychology, University of Alberta \\
2. Neuroscience and Mental Health Institute, University of Alberta \\
3. Alberta Machine Intelligence Institute (Amii) \\
4. Department of Medicine, University of Alberta \\
5. Faculty of Kinesiology, Sport, and Recreation, University of Alberta \\

\vspace{5pt}
*Correspondence: nathan3@ualberta.ca

\end{center}

\vspace{5pt}

\begin{abstract}
\normalsize
\noindent
Progress has led to a detailed understanding of the neural mechanisms that underlie decision making in primates. However, less is known about why such mechanisms are present in the first place. Theory suggests that primate decision making mechanisms, and their resultant behavioral abilities, emerged to maximize reward in the face of noisy, temporally evolving information. To test this theory, we trained an end-to-end deep recurrent neural network using reinforcement learning on a noisy perceptual discrimination task. Networks learned several key abilities of primate-like decision making including trading off speed for accuracy, and flexibly changing their mind in the face of new information. Internal dynamics of these networks suggest that these abilities were supported by similar decision mechanisms as those observed in primate neurophysiological studies. These results provide experimental support for key pressures that gave rise to the primate ability to make flexible decisions.
\end{abstract}

\section*{Introduction}\label{intro}
\addcontentsline{toc}{section}{Introduction}

The process of decision making determines how people choose between entrées at a restaurant, strategies in a competitive game, or votes between political candidates. Decision making in humans and non-human primates is well-studied and its neural mechanisms are increasingly well-understood. Behaviour \citep{ratcliff2008diffusion}, neural recordings \citep{shadlen2001neural}, and causal neural perturbations \citep{salzman1992microstimulation, hanks2006microstimulation} all strongly support the idea that primates make decisions using a common mechanism termed evidence accumulation. Evidence accumulation models state that internal or external information is converted to momentary evidence in support of a decision. Momentary evidence is then accumulated over sequential samples in time to a decision threshold, which determines both choices and response times \citep{gold2007neural}. Extensions to evidence accumulation models are also able to explain complex phenomena such as changes of mind, where primates flexibly change their commitment from one option to an alternative after considering new information \citep{resulaj2009changes, atiya2020changes}.

Theory suggests that primate decision making mechanisms, and their resultant behavioral abilities, emerged via the biological need to act in noisy, temporally uncertain environments \citep{cisek2012making, wispinski2020models}. Recent advances in artificial neural network research afford the ability to experimentally test such theories about emergence—by asking if networks optimized to perform a task given particular constraints and assumptions develop similar properties as the biological systems under investigation \citep{kell2019deep, kanwisher2023using}. For example, deep neural networks have provided compelling accounts for how primate-like image recognition \citep{lindsay2021convolutional, kanwisher2023using} and motion processing \citep{rideaux2020but} emerge when networks are trained to classify natural images via supervised learning. Other work has shown that biological-like mechanisms to solve detour or motor control problems emerge when trained to reproduce behavior from animals \citep{sussillo2015neural, banino2018vector}.

\begin{figure}[!htb]
\centering
\includegraphics[width=0.8\textwidth]{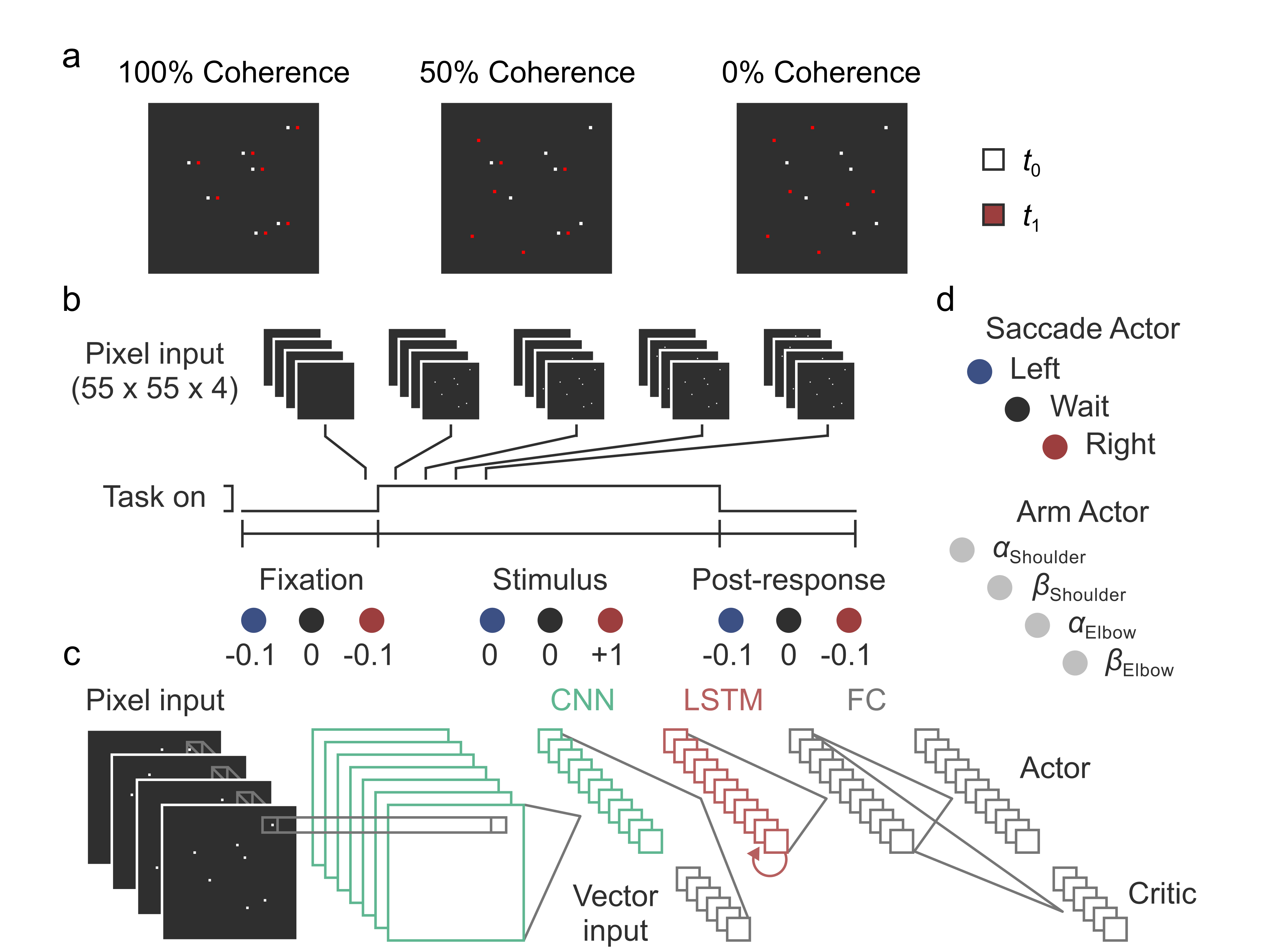}
\caption{
Task and agent network architecture. 
a) Random dot motion discrimination stimuli with varying levels of noise. Dots at $t_{0}$ (white) either move coherently in one direction, or are replaced randomly at $t_{1}$ (red for illustration only) with an independent probability determined by coherence. All examples are from trials where the correct response is ``right”. 
b) Saccadic random dot motion task structure. Agents are rewarded with +1 for responding correctly during the stimulus period, -0.1 for responding before or after the stimulus period, and 0 otherwise. 
c) Actor-critic agent architecture. Pixel and vector input are passed left to right through layers of a deep neural network consisting of convolutional and sum pool operations (CNN), a recurrent layer (LSTM), and fully-connected (FC) layers. Critic output is a single linear unit. Actor output depends on the action space of the task. 
d) Agents responding with saccades have access to discrete ``left”, ``wait”, and ``right” actions. Agents responding with continuous arm movements control shoulder and elbow joint forces, parameterized by alpha ($\alpha$) and beta ($\beta$) parameters for independent beta distributions (see \nameref{methods}). After training, network parameters were frozen and performance was analyzed.
}
\label{fig:fig1}
\end{figure}

Here we ask if primate-like decision making emerges in artificial agents trained to maximize reward in a noisy, temporally uncertain environment. Specifically, we train agents via reinforcement learning to solve the random dot motion discrimination task \citep{shadlen2001neural}. In this task, decision makers are shown dots that move to the left or to the right with some level of random noise (termed coherence), and are asked to report in which direction the dots are moving. Because dot motion is noisy, decision makers need to consider multiple time samples of motion to make an accurate decision. The random dot motion task is widely used in perceptual decision making research in part because it acts as a proxy for an uncertain and dynamic world, and allows for experimental control over environmental noise (Fig \ref{fig:fig1}a, b). We explore two variations in this work: an agent trained to complete the random dot motion task using a simulated saccadic response (as in many non-human primate studies; \citealp{britten1992analysis, roitman2002response}), and an agent trained to complete the random dot motion task by controlling a two-degree-of-freedom arm (similar to collected human data; \citealp{resulaj2009changes, van2016common}; Fig \ref{fig:fig1}c, d).

Below, we identify four key properties of primate decision making that we aim to observe in trained deep reinforcement learning agents to consider them ``primate-like". Here we focus on algorithmic-level properties given the high level of abstraction in modeling primate brains with deep neural networks. As such, we do not consider several other important properties closer to the implementational level of primate decision making such as spike count variance \citep{churchland2011variance}.

First, agents need to display stereotyped behavioral signatures. Animals tend to respond faster, more accurately, and with higher confidence during easy relative to hard decisions across a wide array of decision making tasks \citep{roitman2002response, gold2007neural, wispinski2020models}. These behavioral signatures are sometimes known as the three pillars of choice behavior \citep{shadlen2013decision}. Primate-like agents also need to be able to trade off increases in accuracy at the expense of decision speed; in humans, speed-accuracy trade-offs vary naturally between individuals, but can also be influenced via instructions or reward structures \citep{palmer2005effect, heitz2014speed}.

Second, agent internal dynamics should mirror those found in biological agents. Specifically, agent dynamics should match two functions identified from studies on the neural basis of decision making in primates: learned representations of relevant decision evidence, and the accumulation of this evidence \citep{gold2007neural}. For instance, when making saccadic responses during the random dot motion discrimination task, recordings from primate medial temporal (MT) cortex suggest this area encodes the direction and magnitude of momentary motion on the screen \citep{britten1992analysis}. That is, MT cells selective for motion display tonic firing rates proportional to dot motion direction and coherence. Downstream, the lateral intraparietal area (LIP) is thought to accumulate this momentary motion evidence over time to a decision threshold, which determines in what direction the animal responds with and when \citep{shadlen2001neural, roitman2002response}. Namely, selective LIP cells display a \textit{change} in firing rate proportional to dot motion and coherence. Internal dynamics of primate-like artificial agents should reflect these two distinct properties observed in neurophysiological studies.

Third, agents should display distinguishing characteristics of flexible primate decision making. Specifically, primates are able to change their mind regarding which decision option they prefer in the face of new information—a hallmark of cognitive flexibility \citep{resulaj2009changes, atiya2020changes}. Changes of mind during decision making have been decoded from neural activity in non-human primates \citep{kiani2014dynamics, peixoto2018population}, even in real-time \citep{peixoto2021decoding}. Therefore, artificial agent internal dynamics should suggest the same ability.

Fourth and finally, changes of mind have most notably been observed via movement trajectories in humans \citep{resulaj2009changes, van2016common}, where humans initially reach toward one target before switching in-flight to ultimately choose an alternative target. These changes often correct for initial mistakes, suggesting that they arise from the consideration of additional information to continuously improve accuracy while decision information is available. Changes of mind should not only be inferred from neural dynamics (as in the third key property), but also observed through overt behavior as an agent interacts with the environment.

Below, we investigate each of these four key properties in turn, and show that the trained machine agents meet all of the above criteria for primate-like decision making. We additionally describe two simple changes to agent architecture and training environment in which primate-like decision making does \textit{not} emerge, suggesting critical pressures that contributed to the emergence of flexible decision making in biological agents.

\FloatBarrier
\section*{Results}\label{results}
\addcontentsline{toc}{section}{Results}

\subsection*{Property 1: Agents learn stereotyped decision making behaviors}
\addcontentsline{toc}{subsection}{Agents learn stereotyped decision making behaviors}

\begin{figure}[!htb]
\centering
\includegraphics[width=1.0\textwidth]{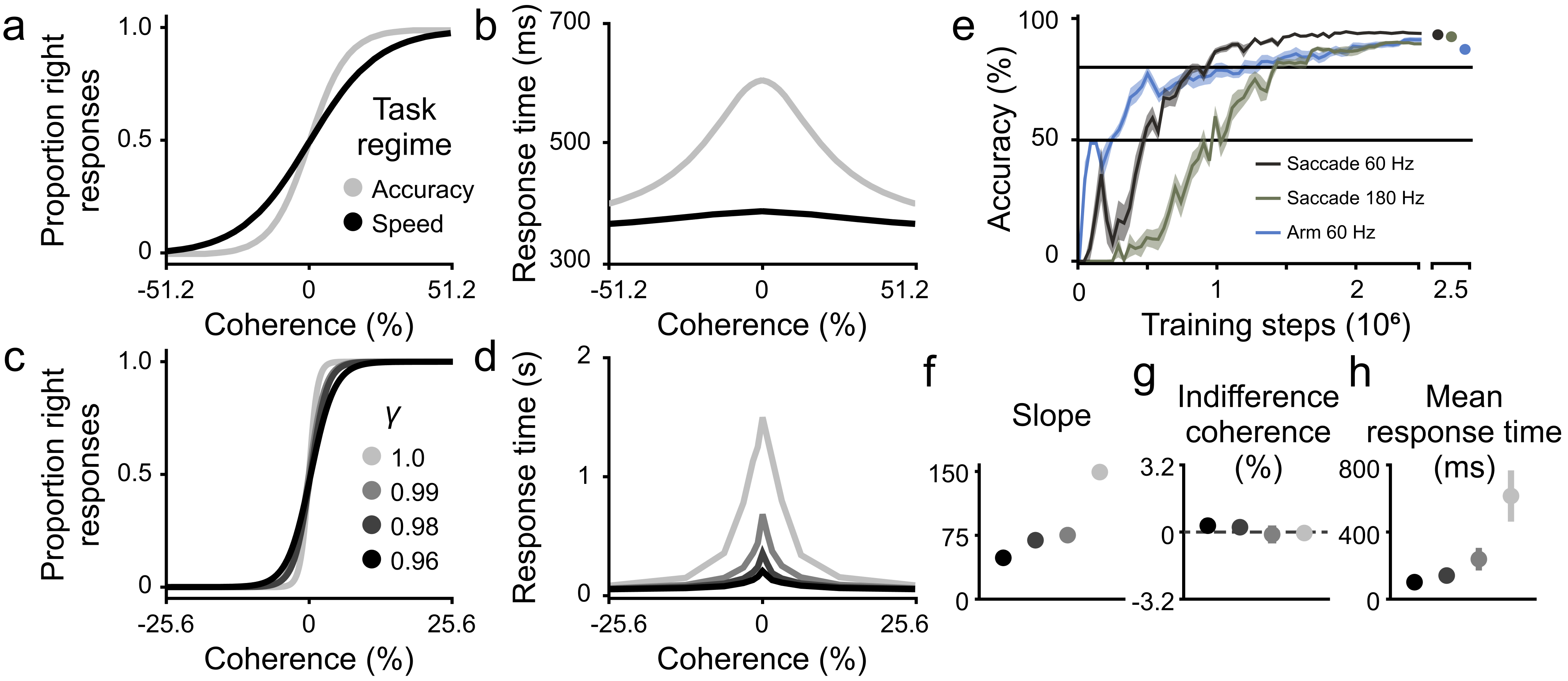}
\vspace{-10pt}
\caption{
Property 1: Agents learn primate-like decision making behaviors.
Agents were trained on the random dot motion task to respond either with discrete actions (left/right saccades), or continuous actions (controlling the shoulder and elbow forces of a simulated arm). Saccade agents were trained at either 60 Hz or 180 Hz, as the 180 Hz agents allow for higher temporal resolution to investigate dynamics (see \nameref{methods}). Results did not differ between 60 and 180 Hz agents.
a) Evidence accumulation model fit to the behavior of a single non-human primate's accuracy and b) response times. Primates completed the random dot motion task and were rewarded with an emphasis on either speed or accuracy \citep{hanks2014neural}.
c) Logistic functions fit to responses grouped by agent discount rate ($\gamma$; N = 15 per $\gamma$). Agents trained with higher discount rates (light gray) were more accurate than those trained with lower discount rates (dark gray). Models are 60 Hz saccade agents.
d) Average response time by agent discount rate. Agents trained with higher discount rates (light gray) were slower to respond than those trained with lower discount rates (dark gray).
e) Periodic evaluation of agents during training (colored traces), and final evaluation after training (dots). Shaded areas (traces) and vertical bars (dots) represent standard errors across 10 random seeds. Horizontal line at 50\% accuracy represents chance performance on the random dot motion task for an agent that always responds within the stimulus period. Horizontal line at 80.3\% represents maximum accuracy achievable from considering only a single time step of dot motion by a hand-constructed evidence accumulation model (see \nameref{methods}).
f) Slopes were extracted from logistic functions fit to each agent’s final evaluation accuracy.
g) Indifference points were extracted from logistic functions fit to each agent's final evaluation accuracy.
h) Mean response time by discount rate condition.
Panels a and b are adapted from non-human primate data in \citet{hanks2014neural}.
}
\label{fig:fig2}
\end{figure}

During perceptual decision making, animals tend to respond faster and more accurately during easy (high absolute coherence) relative to hard (low absolute coherence) decisions \citep[Fig \ref{fig:fig2}a, b;][]{roitman2002response, gold2007neural, wispinski2020models}. Here, trained artificial agents replicated these results with accuracy and response times that similarly varied with dot motion coherence and direction (Fig \ref{fig:fig2}c, d). This pattern of accuracy over coherence levels was well described by a logistic function for all agents ($R^{2} = 0.996 \pm 0.0004$). Trained agents rarely failed to respond by the end of the trial, and successfully learned to withhold responses outside of the dot motion period (mean trials without responses: $0.35\% \pm 0.11\%$).

In the random dot motion task, individual biological decision makers can trade off increases in accuracy at the expense of decision speed \citep{palmer2005effect}. Similarly, by varying a single hyperparameter between artificial agents---the reinforcement learning discount rate ($\gamma$; see \nameref{methods})---we can alter the speed-accuracy tradeoff between individual trained agents at evaluation time. In this case, there was a significant change in the slope (linear mixed effects regression; $b_{1} = 2184.68 \pm 199.21, p = 5.51e\text{-}28$; Fig \ref{fig:fig2}f) of logistic functions fit to choices, showing accuracy increases with $\gamma$. Similarly, mean reaction time increased with $\gamma$ (linear mixed effects regression; $b_{1} = 11335.65 \pm 3162.19, p = 3.47e\text{-}4$; Fig \ref{fig:fig2}h), showing that the speed-accuracy tradeoff is present and controllable in these artificial agents. Finally, no group’s indifference points significantly differed from 0\% coherence (one-sample t-tests, $p\text{s} > 0.05$; Fig \ref{fig:fig2}g), suggesting that changes in $\gamma$ did not bias agents toward the leftward or rightward choice.

The level of accuracy given the time agents took to respond suggests that these agents considered multiple samples of motion in support of their decision. Explicitly simulating an evidence accumulation model shows that agents exceeded the maximum accuracy achievable from considering only a single time step of dot motion, \textit{t}(9) = 33.34, \textit{p} = 4.84e-11 (80.3\% line in Fig \ref{fig:fig2}e; see \nameref{methods}). Accuracy during training suggests that agents started to consider multiple steps of motion in support of their decision after roughly one million steps of experience. In contrast, agents that were instead trained on noiseless motion (i.e., only coherences of 100\%), failed to exceed this one-sample accuracy threshold on average (see \nameref{sup}). Consistent with theory, this suggests that environmental noise was a critical factor in the emergence of primate decision making abilities.

Overall, trained agents displayed key property 1—stereotyped speed and accuracy signatures consistent with primate-like decision making. However, patterns of response speed and accuracy alone are not enough to support claims of primate-like decision making. It is possible for primate-like speed and accuracy patterns to arise from non-primate-like decision making mechanisms \citep{stine2020differentiating, hyafil2023temporal}. For example, an extrema detection mechanism compares individual samples of evidence against response thresholds \textit{without} accumulation. Under this mechanism, a decision maker waits until an individual sample is large enough to trigger a response. This is in contrast to mechanisms that \textit{do} accumulate evidence toward a threshold, which have significantly more empirical evidence in primate neurophysiological experiments \citep{stine2020differentiating}. Therefore, to distinguish between primate-like and non-primate-like decision mechanisms, we now turn to our second key property—the internal dynamics of these trained agents.

\FloatBarrier
\subsection*{Property 2: Dynamics reflect momentary and accumulated decision evidence}
\addcontentsline{toc}{subsection}{Dynamics reflect momentary and accumulated decision evidence}

To support a claim of primate-like decision making, agent internal dynamics should be selective for both momentary and accumulated motion direction and coherence, consistent with an evidence accumulation framework and neuronal recordings from primate areas MT and LIP \citep{britten1992analysis, shadlen2001neural, roitman2002response}. Here we investigate individual unit activations from the sum pool layer of the network (CNN; Fig \ref{fig:fig1}b), and the recurrent layer of the network (LSTM; Fig \ref{fig:fig1}b) as rough analogues for areas MT and LIP in primates during saccadic decision making (see \nameref{methods}).

\begin{figure}[!htb]
\centering
\includegraphics[width=1.0\textwidth]{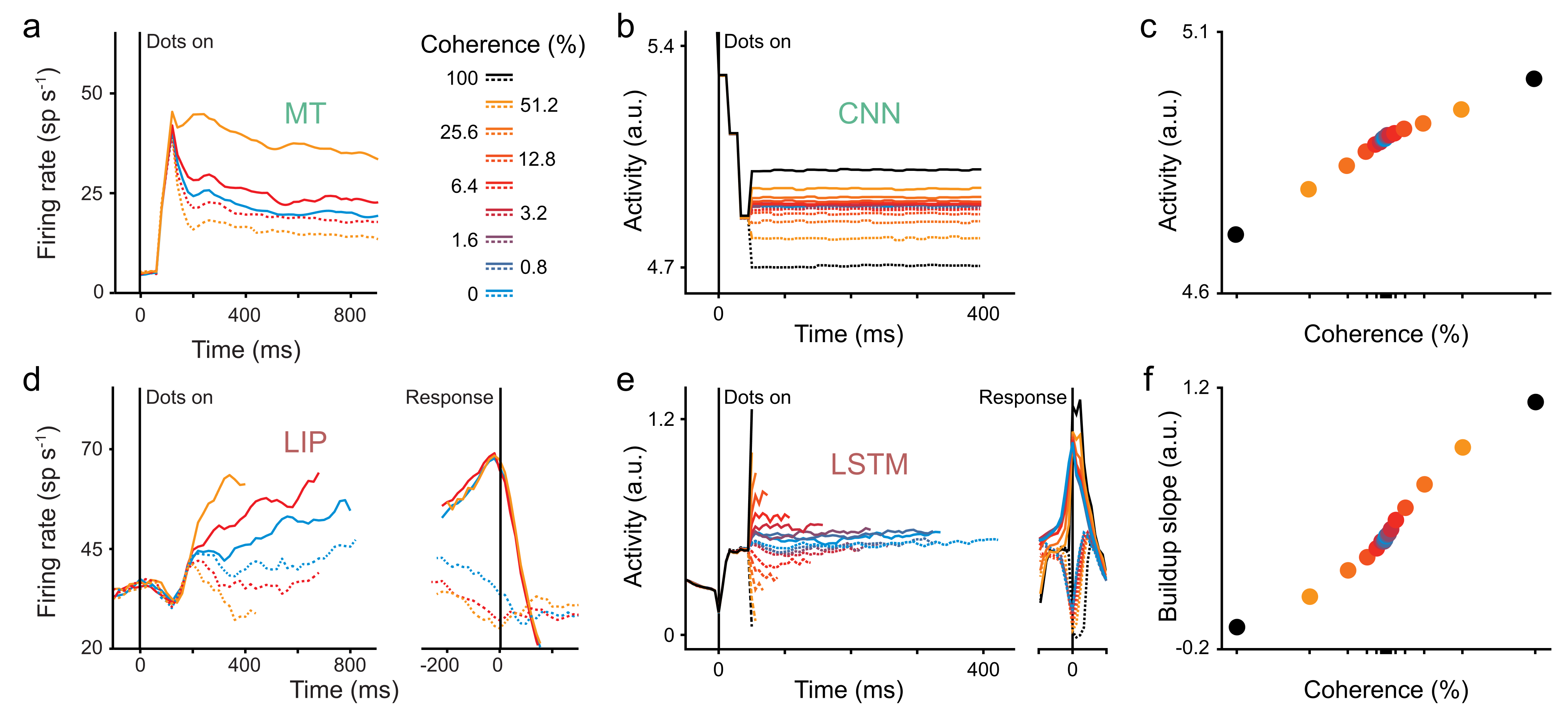}
\vspace{-20pt}
\caption{
Property 2: Agents learn primate-like internal decision making dynamics.
Internal dynamics of a representative agent in the 180 Hz saccade task. Solid lines indicate rightward motion trials, and dashed lines indicate leftward motion trials. Data are from correct trials only. For dynamics on the left, individual trials making up the plotted averages are considered up until the first of: response time or median response time.
a) Average response of direction selective MT neurons recorded in a primate during the random dot motion task \citep[adapted from][]{gold2007neural}.
b) An example CNN unit with average activity proportional or inversely proportional to momentary motion coherence and direction, similar to primate area MT.
c) The average activity of this CNN unit is roughly linearly related to motion coherence.
d) Average response of LIP neurons recorded in a primate during the random dot motion task \citep[adapted from][]{gold2007neural}. On the left, neural activity is time-locked to dot motion onset. On the right, activity is time-locked to primate saccadic response.
e) An example LSTM unit with activity proportional or inversely proportional to accumulated momentary motion coherence and direction, similar to primate area LIP.
f) The average buildup slope, from dot motion onset to the first step of full motion, of this LSTM unit is roughly linearly related to coherence.
}
\label{fig:fig3}
\end{figure}

Several CNN units ($75\% \pm 1.4\%$) consistently showed a sustained response proportional or inversely proportional to momentary motion strength and direction (Fig \ref{fig:fig3}b, c; see \nameref{sup}). These results are consistent with the idea that CNN units allow the agent to compute a measure related to momentary motion evidence on every time step, similar to the proposed function of primate area MT in this task \citep[Fig \ref{fig:fig3}a;][]{gold2007neural}. In addition, several LSTM units ($72.9\% \pm 1.3\%$) revealed accumulation dynamics (Fig \ref{fig:fig3}e). That is, they showed a significant relationship between buildup \textit{slope} and coherence (Fig \ref{fig:fig3}f; see \nameref{sup}), as in primate area LIP \citep[Fig \ref{fig:fig3}d;][]{roitman2002response, o2018bridging}. Further, when LSTM dynamics were aligned to response, the activity of a subset of units appeared to meet or exceed a threshold level, mimicking the proposed decision threshold gating saccadic responses \citep{gold2007neural}. Specifically, these response-aligned dynamics were most consistent with a decision threshold that collapsed over time \citep[e.g.,][]{drugowitsch2012cost}. LSTM and CNN dynamics were both robust to the training of several agents that differed only in initial random seeds. Together, these dynamics suggest agents learned a two-part evidence accumulation mechanism, which processed momentary motion energy from raw pixels, and then accumulated this momentary motion energy to a decision threshold. Such dynamics satisfy our second key property of primate-like decision making.

In contrast, if agents instead learned an alternative, non-accumulation mechanism such as extrema detection \citep{stine2020differentiating}, units selective for accumulated decision evidence would not be predicted to emerge. Agents trained without recurrence showed several units selective for momentary decision evidence (see \nameref{sup}). However, downstream dynamics strongly suggest that these non-recurrent agents make non-primate-like decisions, based on a single, extreme sample of motion (see Fig \ref{fig:ablations}). These results suggest that recurrence---often taken for granted in biological systems, but less prominent in many modern deep learning systems---is a second key ingredient for the emergence of primate-like decision making.

\FloatBarrier
\subsection*{Property 3: Changes of mind decoded from dynamics}
\addcontentsline{toc}{subsection}{Changes of mind decoded from dynamics}

We next look at the adaptive behavior of these trained agents by investigating changes of mind—a phenomenon where a decision maker revises their decision online in the face of new information \citep{resulaj2009changes}. Changes of mind (CoMs) have been decoded from neural activity in primates \citep{kiani2014dynamics, peixoto2018population}, even in real-time \citep{peixoto2021decoding}, and have their own key properties \citep{peixoto2021decoding}. First, changes of mind are more frequent when decisions are more difficult (i.e., during low relative to high coherence trials). Second, changes of mind are more likely to be corrective than erroneous. In other words, changes more likely move from an initially incorrect choice to an ultimately correct one rather than vice versa, indicating that these changes are based on additional information during the decision process to improve accuracy.

\begin{figure}[!htb]
\centering
\includegraphics[width=1.0\textwidth]{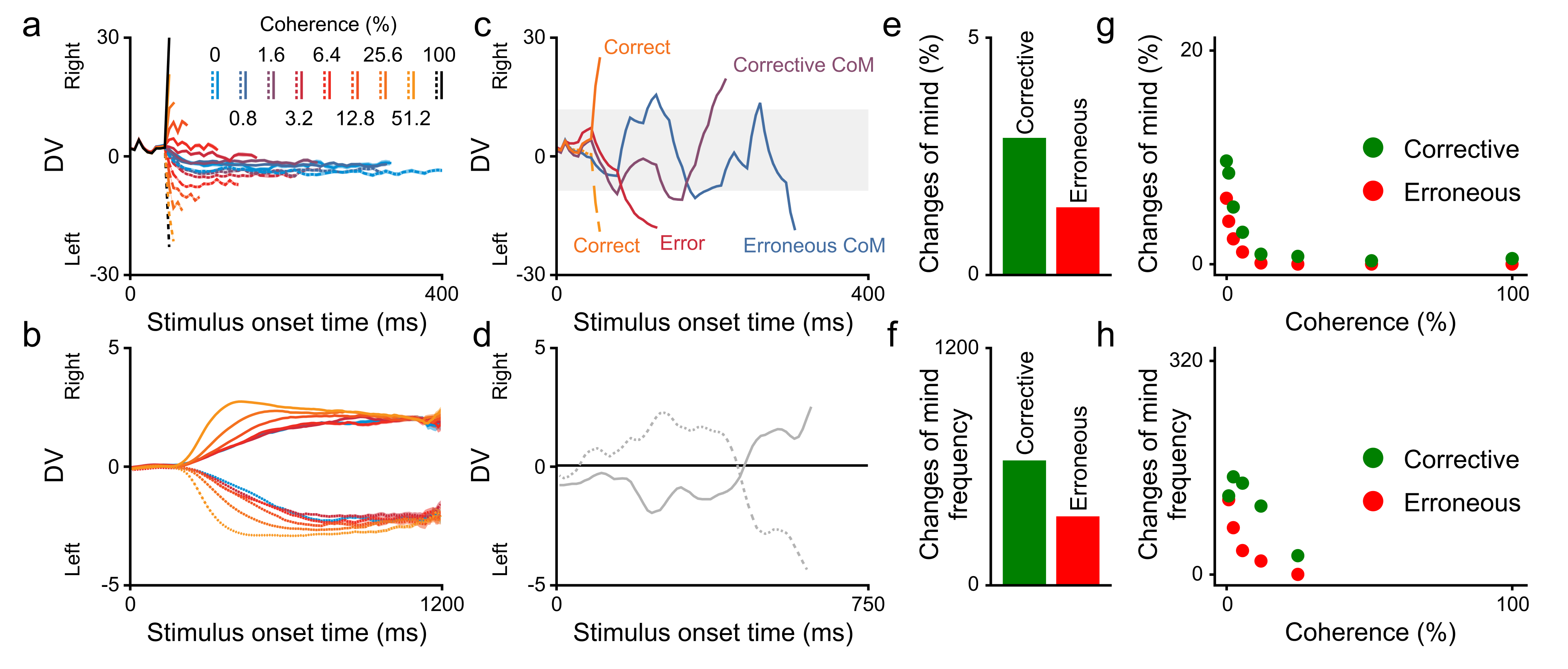}
\caption{
Property 3: Decoding of neural changes of mind.
a) Logistic decoder decision variable (DV) by coherence condition from a representative agent trained in the 180 Hz saccadic task. Shaded regions denote standard errors. Traces plotted until median response time. Solid traces indicate rightward motion trials, and dashed traces indicate leftward motion trials.
b) Logistic decoder DV by coherence condition pooled across two non-human primates. Traces are plotted for the full time, as the primate needed to wait for a decision cue before response \citep{peixoto2021decoding}.
c) Example of single-trial DV traces. Line colors correspond to the coherence conditions of each single trial. Shaded region denotes change of mind threshold (see \nameref{methods}).
d) Example of two single-trial DV traces decoded from a non-human primate that were classified as change of mind trials.
e) Decoded changes of mind. Agents displayed more corrective (green) than erroneous (red) changes of mind.
f) Decoded changes of mind frequency from a non-human primate.
g) Decoded changes of mind by coherence. Corrective changes (green) are those where the model DV at some point predicts an incorrect choice before an ultimately correct choice is made. Erroneous changes (red) are those where the model DV predicts a correct choice before an ultimately incorrect choice is made.
h) Frequency of decoded changes of mind by coherence in a non-human primate.
Bottom panels (b, d, f, and h) are adapted from non-human primate data in \citet{peixoto2021decoding}.
}
\label{fig:fig4}
\end{figure}

Similar to primate studies \citep{kiani2014dynamics, peixoto2018population, peixoto2021decoding}, we train a linear decoder to predict the choices of a decision maker throughout a trial based on evolving neural activity. Specifically, we train a logistic regression classifier, which decodes a decision variable (DV) to predict left/right choices based on LSTM layer activity at every time step. Changes in the decoder’s prediction (i.e., a change of sign in the DV) before a response suggest a neural change of mind (Fig \ref{fig:fig4}a, b, c, d).

Consistent with non-human primate experiments, changes of mind were more frequent when decisions were more difficult (Fig \ref{fig:fig4}g, h; linear regression of the proportion of CoMs by log coherence; $b_{1} = -0.33 \pm 0.006, p = 0.001$). Importantly, changes of mind were more likely to be corrective than erroneous (Fig \ref{fig:fig4}e, f; chi-squared goodness-of-fit test; $\chi^{2} = 43.28, p = 4.74e\text{-}11$). Overall, these results show that the trained artificial agents considered here are able to learn highly flexible, error-correcting behavior similar to primates.

\FloatBarrier
\subsection*{Property 4: Changes of mind in movements}
\addcontentsline{toc}{subsection}{Changes of mind in movements}

Ballistic tasks such as those requiring a saccadic response are the exception rather than the rule of real world decision making in animals. More typically, animals must execute temporally and spatially extended movements to interact with the world \citep{wispinski2020models}. When biological agents move to make a decision, valuable time elapses, and the consideration of decision information extends into movement \citep{resulaj2009changes, wispinski2020models, michalski2020reaching}. This continuous consideration of information throughout a movement leads to the final key property of primate-like decision making we consider—decision-related movement fluctuations \citep{song2009hidden, wispinski2020models}, and changes of mind observable in movement trajectories \citep{resulaj2009changes, van2016common}.

We trained a new set of end-to-end agents that instead responded by controlling the joint forces of a simulated two-degree-of-freedom planar arm to move a fingertip toward a left or a right target (see \nameref{methods}). We compared this agent behavior to collected data from 13 humans in a similar reaching task, where participants would perform the random dot motion discrimination task by reaching to one of two targets on an upright screen. Human participants also rated their confidence after every decision, and feedback was withheld.

\begin{figure}[!htb]
\centering
\includegraphics[width=0.6\textwidth]{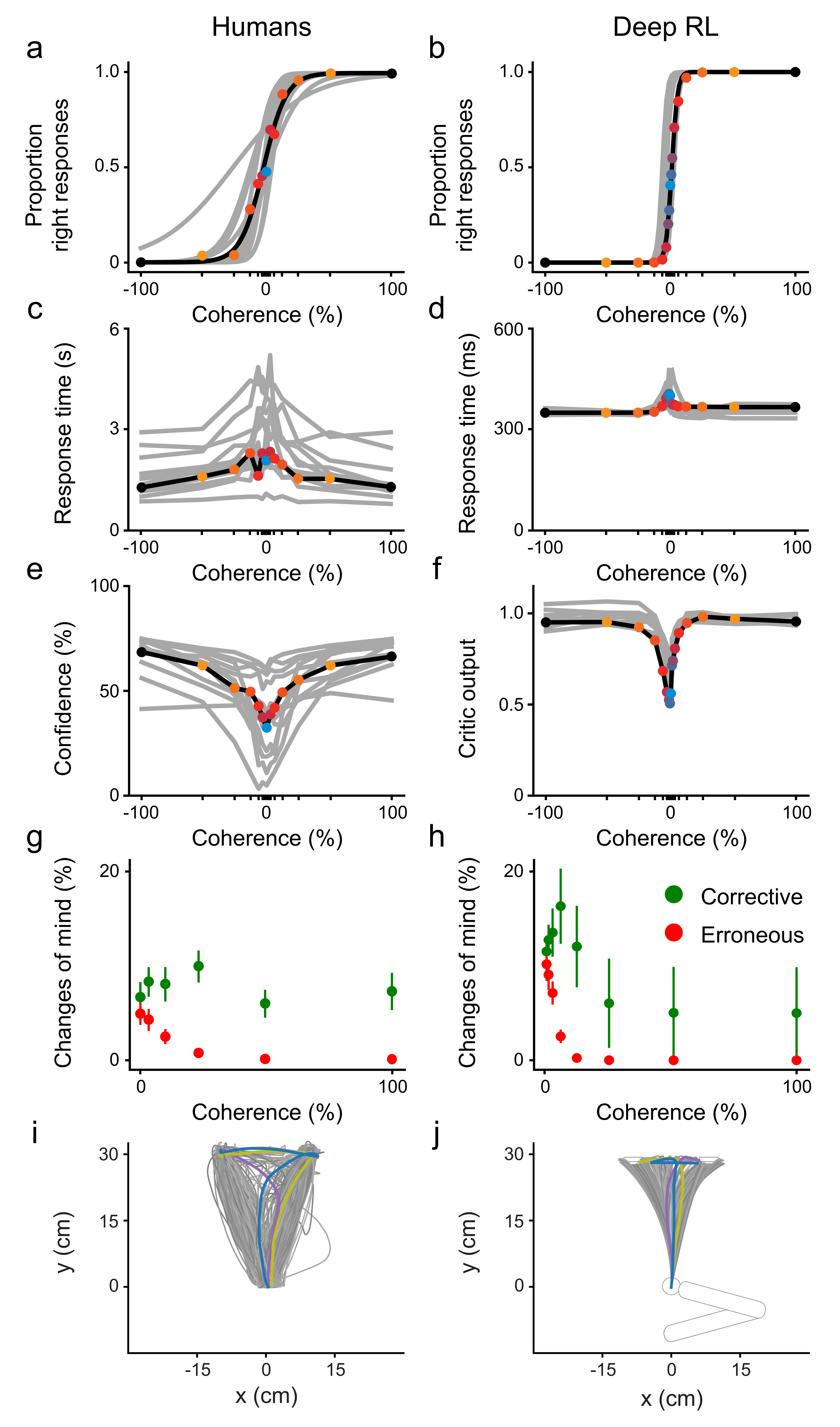}
\caption{
    Property 4: Behavioral changes of mind. 
    Comparison between human reaching performance (left column; N = 13) and trained deep reinforcement learning agents controlling a 2DOF arm (right column; N = 10). Individual representative human and trained agent highlighted (black line, coloured dots), and all other individuals (gray lines).
    a, b) Accuracy. Logistic function fits; humans: $R^{2} = 0.957 \pm 0.019$; artificial trained agents: $R^{2} = 0.998 \pm 0.001$.
    c, d) Response time. Note differences in scaling between humans and artificial agents.
    e, f) Human post-decision confidence ratings, and network critic output at response time.
    g, h) Mean changes of mind across all subjects. Vertical lines indicate standard errors. i, j) Movement trajectories from an individual representative human and artificial agent (2DOF arm shown), with selected changes of mind highlighted.
}
\label{fig:fig5}
\end{figure}

In the above sections considering the saccade agents, decoding an agent’s internal state reveals the evolution of an agent’s decision between left and right options throughout a trial (Fig \ref{fig:fig4}). In this continuous control task, decision states can instead be inferred from a decision maker’s movements in physical space between left and right options. Looking at these movements, both humans and artificial agents displayed more curved movement trajectories on hard trials, and changed their mind while moving to correct for initial errors (Fig \ref{fig:fig5}i, j). Similar to changes of mind decoded from neural activity, behavioral changes of mind were more frequent when decisions were more difficult (Fig \ref{fig:fig5}g, h; linear mixed effects regression; Humans: $b_{1} = -0.055 \pm 0.017, p = 0.0015$; Artificial agents: $b_{1} = -0.20 \pm 0.087, p = 0.024$). Importantly, these behavioral changes of mind tended to be more corrective than erroneous (paired one-tailed t-tests; Humans: $t(12) = 6.11, p = 2.61e\text{-}5$; Artificial agents: $t(9) = 1.86, p = 0.048$), suggesting that both the humans and artificial agents flexibly altered movements online as a result of incoming information. In addition to behavioral similarities, the trained agents in this continuous control task developed similar internal dynamics as the trained agents in the saccadic response task, suggesting that these agents also continuously integrate evidence over time in support of their decision (see \nameref{sup}).

Up to now, we have only considered speed and accuracy behavior (Fig \ref{fig:fig5}a, b, c, d), but not the third pillar of choice behavior—confidence \citep{shadlen2013decision}. Animals tend to respond with higher confidence during easy relative to hard decisions across a wide array of tasks. Results show human post-decision confidence judgements follow these established patterns (Fig \ref{fig:fig5}e). However, decision confidence is difficult to determine without language. For example, without the ability to directly ask about the confidence of non-human primate decision makers, researchers employ alternative tasks such as post-decision wagering \citep{kepecs2014computational}, or analyze neural patterns \citep{kepecs2008neural}.

While we cannot ask the current reinforcement learning agents to verbalize metacognitive judgments about their confidence after each decision, we can leverage their neural architecture to answer similar questions. The current agents use an actor-critic reinforcement learning method \citep[see \nameref{methods};][]{sutton2018reinforcement}. In short, part of the agent’s output represents its policy—the probabilities of each action to be selected on each time step (i.e., the actor; Fig \ref{fig:fig1}b). The other part of the agent’s output represents the agent’s estimate about future cumulative discounted rewards (i.e., the critic; Fig \ref{fig:fig1}b). In brief, actor-critic methods in reinforcement learning work by improving the critic’s estimate about rewards based on experiences, and changing the probabilities of actions relative to the critic’s estimates \citep{sutton2018reinforcement}. Here we can query the critic’s reward estimate on the final step of each trial, just before the artificial agent touches one of the two targets. This value output approximates the agent’s prediction of a correct response (+1) over an incorrect response (0; see \nameref{methods}). Looking at these values, the pattern of agent critic output emerged to closely match the pattern of human post-decision confidence judgements (Fig \ref{fig:fig5}f)—both consistent with theory \citep{wispinski2020models} and primate experiments \citep{kiani2009representation}. Overall, these patterns show a proxy for emergent primate-like confidence in artificial agents, completing the three pillars of choice behavior \citep{shadlen2013decision}.

\FloatBarrier
\section*{Discussion}\label{discussion}
\addcontentsline{toc}{section}{Discussion}

Here we show that artificial agents trained to maximize reward in the face of noisy, temporally evolving information learn behaviors and internal dynamics similar to primate decision makers. Agents learned to trade off speed and accuracy, to estimate future rewards which closely aligned with human post-decision confidence judgements, and to flexibly change their mind in the face of new information. These results also suggest an underlying decision making mechanism similar to evidence accumulation, which is thought to underlie many perceptual- and value-based decisions in biological agents \citep{shadlen2016decision}. Evidence accumulation mechanisms have previously emerged in supervised learning \citep{mante2013context} and reinforcement learning \citep{song2017reward} models given noisy numerical input. The models shown here extend this work both upstream and downstream by learning adaptive decision making policies end-to-end---from raw pixels to continuous effector output.

That these phenomena presented here emerge in artificial agents support theory that these mechanisms emerged similarly in biological systems—via the need to act in noisy, temporally uncertain environments \citep{cisek2012making, wispinski2020models}. It may seem unsurprising that the agents considered here discover a similar mechanism to evidence accumulation, given that evidence accumulation models are thought to be the optimal solution to several perceptual decision making tasks \citep{ratcliff2016diffusion}. However, we identify two simple changes where primate-like decision making does \textit{not} emerge. Agents without recurrence, and agents trained on noiseless motion were both able to complete the task. However, both failed to reliably display key properties of primate-like decision making, suggesting at least two critical pressures for primate-like decision making to emerge—recurrence and environmental noise.

The current results fit with other work showing that training artificial agents on biological tasks encourages biological-like solutions, for example motion processing \citep{rideaux2020but}, detour problems \citep{banino2018vector}, and patch foraging \citep{wispinski2022adaptive}. Overall, these results support long-standing theory that biological intelligence (and biological-like artificial intelligence) may emerge through optimization in environments that mirror developmental \citep{turing1950mind, smith2005development} or evolutionary pressures \citep{cisek2019resynthesizing} faced by biological systems \citep{kell2019deep, kanwisher2023using}.

The agents considered here are simulated at a high level of biological abstraction and do not consider several components relevant to biological researchers. Future work may further consider the simulation of decision making using spiking neural networks \citep{lillicrap2016random}, biologically plausible methods of network weight updating \citep{bengio2016feedforward}, or considering architectures with long-range recurrence. The current agents are also simulated with no neural or motor delays or noise, which is apparent when looking at the low response times and low behavioral and neural variability of these agents relative to biological decision makers. Future work may consider integrating these limitations into artificial agents to better understand biological solutions via deep reinforcement learning.

Overall, we present artificial agents that learn via rewards to enact flexible decisions in a primate-like way from raw sensory information. We argue these results provide unique insight into the origins of primate decision mechanisms, and provide potential applications for automating human decision making in noisy, changing environments.

\newpage

{\small
\paragraph{Acknowledgments.}
This research was supported by the National Science and Engineering Research Council of Canada (NSERC) and the Killam Trusts. This research was enabled in part by computational support provided by the BC and the Prairie DRI Groups, and the Digital Research Alliance of Canada (\texttt{alliancecan.ca}). Work by P.M.P. was supported in part by the Canada CIFAR AI Chairs program. We would like to thank Garett Motley for assisting in human data collection.

\paragraph{Authors' contributions.}
All authors contributed to idea generation and manuscript writing.
N.J.W. carried out human and computational experiments and data analysis.
Human experiments were programmed by N.J.W. and C.S.C. Computational experiments were programmed by N.J.W. and S.A.S.

\paragraph{Conflict of interest.}
The authors declare no competing interests.

\paragraph{Availability of data and materials.}\label{datalinks}
Videos, data, and code for agent training and analysis will be made publicly available at the following websites upon publication: \\
\texttt{https://osf.io/sdhpg/} \\
\texttt{https://github.com/nathanwispinski/primate-like-decisions}
}

\setlength{\bibsep}{2.5pt plus 1.0ex}  

\addcontentsline{toc}{section}{References}
\bibliographystyle{apacite}
\bibliography{sample}

\newpage

\section*{Methods}\label{methods}
\addcontentsline{toc}{section}{Methods}

\subsection*{Motion discrimination task}
\addcontentsline{toc}{subsection}{Motion discrimination task}

Agents completed a reaction time version of the random dot motion discrimination task commonly used in studies of decision making in humans and non-human primates \citep{shadlen2001neural, roitman2002response}. Agents were shown a video of noisy dots moving to the left or to the right, and guessed in which direction the dots were moving and when to respond.

On each frame, seven dots were drawn within a circular aperture within a 55 x 55 pixel image. Each dot had an independent probability of moving at a fixed angle and distance from its current location on the next frame, or being drawn at a new random location. This probability of dot motion, also known as coherence, determined the strength of motion on each trial. Although the expectation of motion strength in each coherence condition over time is constant, motion strength is noisy on single trials. During training, dots could move either left or right, and dot motion strength on each trial was selected from seven coherence values used in primate decision research: [0\%, 3.2\%, 6.4\%, 12.8\%, 25.6\%, 51.2\%, 100\%]. Motion direction and coherence varied randomly from trial-to-trial, but remained fixed within each trial. Responses on zero coherence trials were rewarded with 50\% probability since there was no correct response on these trials. The dot motion stimulus was simulated at 60 Hz, in line with refresh rates of motion stimuli presented to mammals \citep{shadlen2001neural, katz2016dissociated}.

Dot motion discrimination stimuli are often presented with three interleaved sets of motion, such that dots on frame one are moved or randomly replaced on frame four, while an independent set of dots are presented on frames two and five, etc. \citep{shadlen2001neural}. During training, agents were presented either dot motion stimuli with three interleaved sets of motion (as in a typical biological experiment), or stimuli with no interleaved frames to approximate natural consistent motion \citep[as in][]{rideaux2020but}, randomly determined at the start of each training trial with equal probability. After training, evaluation and all subsequent analyses were completed using motion stimuli with three interleaved frames. Dot speed was kept the same regardless of the number of interleaved frames—in other words, coherently moving dots were displaced one pixel horizontally for stimuli with no interleaved frames, and three pixels horizontally for stimuli with three interleaved frames (see \nameref{datalinks} for videos).

Agents responded via a simulated saccadic response as in most non-human primate decision research, or a simulated reaching movement to one of two targets modeled after human data (see \nameref{humandata}). In the saccadic task, agents responded with discrete ``left”, ``right”, or ``wait” actions. In the reaching task, agents controlled the shoulder and elbow forces (continuous, [-1, +1]) of a two-degree-of-freedom planar arm in the MuJoCo physics engine \citep{todorov2012mujoco}.

Correct responses during dot motion presentation corresponded to a reward of +1, while incorrect responses corresponded to a reward of 0. If an agent responded before stimulus onset or after stimulus offset, it received a reward of -0.1. ‘Wait' actions on every time step before, during, and after stimulus presentation corresponded to a reward of 0. In the reaching version of the task, agents were additionally rewarded on each time step based on the forward distance (in meters) of the simulated fingertip (multiplied by a scaling coefficient) to encourage forward reaching movements. Specifically, the fingertip of the agent started at a distance of 0 m on every trial, and the two targets and screen were located 0.3 m forward from this start position. With a scaling coefficient of 0.005 (see Supplementary Table \ref{hyperparam-table}), this meant that agents would receive an additional reward of 0 on every step at the start position, and a maximum additional reward of 0.0015 on every time step when the finger was touching the screen. Agents with this small reward for moving forward learned the task quicker and more reliably than those without the reward, as it encouraged the agents to explore states further from the start position early on in training.

The saccadic random dots task was simulated as trials that ended 5 steps after a response, or after 3 seconds without a response. For the saccadic agent simulated at 180 Hz, the trial ended 15 steps after a response, or after 2 seconds without a response. For the reaching task, trials always ended when the fingertip made contact with the target, or after 3 seconds without a response. Stimuli onset times on each trial were drawn from a random uniform distribution (see Supplementary Table \ref{hyperparam-table}; Fig \ref{fig:fig1}b), and dot motion was extinguished immediately after a response for the remainder of the trial. For the reaching task, the agent’s arm was held in place until 4 frames of dot motion had been input to the network.

Saccadic agents were simulated at both 60 Hz and at 180 Hz. 60 Hz agents performed one forward pass of information through the neural network and selected actions in sync with new dot motion frames. In contrast, 180 Hz agents performed three forward passes, and corresponding action selections, per new dot motion frame. While slower and more difficult to train, 180 Hz agents allow for investigation of internal dynamics at a higher temporal resolution than 60 Hz agents. Results did not qualitatively differ between 60 Hz and 180 Hz agents.

\subsection*{Network architecture}
\addcontentsline{toc}{subsection}{Network architecture}

All networks were implemented in Python version 3.9 using The DeepMind JAX Ecosystem \citep{deepmind2020jax}.

\qquad The neural network described below accepted 55 x 55 x 4 pixel input, corresponding to the most recent four frames of the dot motion stimulus. Stacked frames as input have been used in deep reinforcement learning agents for playing Atari games \citep{mnih2015human}, and in supervised learning networks to recreate several properties of primate motion-selective area MT \citep{rideaux2020but}.

At each time step, input of shape 55 x 55 x 4 was convolved with 64 3D kernels each with a shape of 5 x 5 x 4 to produce 64 convolutional output maps. Input was padded with zeros so that each convolutional output map was of shape 55 x 55. Units used a rectified linear (ReLU) activation function to model neurophysiological data \citep{rideaux2020but}. Each convolutional output map was then summed so that network output was reduced to 64 values (one for each map). Maps were summed across space as the dot motion discrimination task relies on global, rather than local, perception of motion. Convolution as a first operation was chosen because of the sharing of parameters across image space as in many deep learning models of biological vision \citep{lindsay2021convolutional}, and because the convolution of stacked frames of moving images has been shown to approximate key properties observed in primate area MT \citep{rideaux2020but}. These 64 sum-pooled outputs are referred to as ``CNN” throughout.

The 64 CNN outputs were then concatenated with a vector of task-relevant inputs. In the saccadic version of the task, these were a binary task off signal, a binary task on signal, the agent’s action on the previous step, and the reward on the previous step. In the reaching version of the task, vector inputs additionally included the sine and cosine of shoulder and elbow angles, the velocity of the shoulder and elbow, and the x and y distances of the fingertip to both left and right targets.

After concatenation, inputs were fed to a layer of 128 LSTM units \citep{hochreiter1997long}. LSTM units introduce recurrence by copying their internal ‘cell state’ between time steps. LSTM units are also gated by ‘forget’, ‘input’, and ‘output’ gates, which allow these units to choose to forget information, allow new information to enter, and contextually output memory contents at each time step. The dynamics of these units are governed by the standard equations:

\vspace{-10pt}
\begin{gather}
    i_t = \sigma(W_{ii} x_t + W_{hi} h_{t-1} + b_i) \nonumber \\
    f_t = \sigma(W_{if} x_t + W_{hf} h_{t-1} + b_f) \nonumber \\
    g_t = \tanh(W_{ig} x_t + W_{hg} h_{t-1} + b_g) \nonumber \\
    o_t = \sigma(W_{io} x_t + W_{ho} h_{t-1} + b_o) \nonumber \\
    c_t = f_t \cdot c_{t-1} + i_t g_t \nonumber \\
    h_t = o_t \cdot \tanh(c_t) \nonumber
\end{gather}

Where $x_{t}$ is the LSTM input at time $t$, $h_{t}$ is the hidden state, $i_{t}$, $f_{t}$ and $o_{t}$ are the input, forget, and output gate activations, $c_{t}$ is the cell state, $g_{t}$ is a vector of cell state updates, and $\sigma$ is the sigmoid function. All LSTM states were initialized to zero at the beginning of each trial. Recurrent units, such as LSTMs, were chosen because they have been shown in cases to approximate the accumulation of evidence in favor of a decision given noisy numerical input \citep{song2017reward}. LSTM layer output was finally fed through a fully-connected layer with 128 ReLU units.

Output was then fully-connected to independent actor and critic network heads, consistent with many deep reinforcement learning architectures \citep{arulkumaran2017deep}. The critic network head consisted of 64 fully-connected units with ReLU activations connected to a single linear unit. The critic head acts to estimate the expected return of the agent’s policy given the current state, $s$ (see \nameref{training}).

In the saccadic network, the actor network head consisted of 64 fully-connected units with ReLU activations connected to three linear output units, one for each action available at every time step (left, wait, right). A softmax operation was performed on the three output units so that their sum was equal to one, and an action at each time step was randomly selected from these probabilities. This part of the network determined the agent’s action policy, $\pi$.

For the reaching network, the final layer of the actor head consisted of four output units with a shifted softplus activation function ([1, $\infty$]). These outputs corresponded to alpha and beta parameters for a beta distribution \citep{chou2017improving}—one for shoulder and one for elbow forces. During training, joint forces were randomly sampled from beta distributions parameterised by the network at each time step.

\subsection*{Training}\label{training}
\addcontentsline{toc}{subsection}{Training}

Reinforcement learning was implemented by the Proximal Policy Optimization (PPO) algorithm \citep{schulman2017proximal}. In brief, the full objective function is a weighted sum of a clipped policy gradient loss ($L_{\pi}$), a state-value function loss ($L_{V}$), and an entropy regularization term ($L_{H}$):

\vspace{-10pt}
\begin{gather}
L = L_{\pi} + \beta_{V} L_{V} + \beta_{H} L_{H} \nonumber
\end{gather}

Where $\beta_{V}$ and $\beta_{H}$ are coefficients (see Supplementary Table \ref{hyperparam-table}). The clipped policy gradient loss ($L_{\pi}$) is defined by:

\vspace{-10pt}
\begin{gather}
L_{\pi}(\theta) = \hat{E}_{t}[\text{min}(\rho_{t}(\theta)\hat{A}_{t}, \text{clip}(\rho_{t}(\theta), 1-\epsilon, 1+\epsilon)\hat{A}_{t})]  \nonumber \\
\rho_{t} = \frac{\pi_{\theta}(a_{t} \mid s_{t})}{\pi_{\theta_{\text{old}}}(a_{t} \mid s_{t})}  \nonumber
\end{gather}

Where $\hat{E}_{t}$ denotes the empirical expectation over time steps, $\rho_{t}$ is the ratio of state-action probabilities under the new and old policies, respectively, and $\epsilon$ is a clipping hyperparameter. $\hat{A}_{t}$ is the estimated advantage at time $t$, defined by the truncated generalized advantage estimator from \citet{schulman2017proximal}:

\vspace{-10pt}
\begin{gather}
\hat{A}_{t} = \delta_{t} + (\gamma \lambda)\delta_{t+1} + ... + (\gamma \lambda)^{T-t+1}\delta_{T-1} \nonumber \\
\delta_{t} = r_{t} + \gamma V_{\theta}(s_{t+1}) - V_{\theta}(s_{t}) \nonumber
\end{gather}

Where $\gamma$ is a discount factor $\in$ [0, 1], $\lambda$ is a mixing parameter $\in$ [0, 1], $r_{t}$ is the experienced reward at time $t$, $V(s_{t})$ is the critic head output given state $s_{t}$, and $t$ specifies the time index in [0, T] within a given length-T trajectory segment. The state-value function loss ($L_{V}$) is defined by \citep{schulman2017proximal, huang2022cleanrl}:

\vspace{-10pt}
\begin{gather}
L_{V}(\theta) = \hat{E}_t[(R_{t} - V_{\theta}(s_{t}))^{2}] \nonumber
\end{gather}

Where $R_{t}$ is the TD($\lambda$) return \citep{sutton2018reinforcement}. Finally, the entropy regularization term is the mean entropy of the policy distribution given all states in the batch:

\vspace{-10pt}
\begin{gather}
L_{H}(\theta) = \hat{E}_t[H(\pi_{\theta}(\cdot \mid s_{t}))] \nonumber
\end{gather}

Network parameters were updated after collecting each batch of experiences using backpropagation through time and the Adam method for gradient-based optimization \citep[][see Supplementary Table \ref{hyperparam-table}]{kingma2014adam}. Unless stated, all agents were run with the same hyperparameters and different random seeds initializing network parameters and the starting environment state. Hyperparameters were chosen through an informal search.

All 10 random seeds of the 60 Hz saccade agents, 180 Hz saccade agents, and 60 Hz reaching agents exceeded 80.3\% accuracy at final evaluation (see \nameref{EvAccMethods}), and so were not rejected from analysis. Distributed training used 20 parallel CPU cores, and took approximately 10, 20, and 12 hours for the 60 Hz saccade, 180 Hz saccade, and 60 Hz reaching agents, respectively (including online and post-training evaluation).

After training, network parameters were frozen and agents were evaluated on a version of the task with three interleaved sets of motion. 500 trials were simulated in each coherence and direction condition. Consistent with many deep reinforcement learning experiments \citep[e.g.,][]{lillicrap2016random}, agents were evaluated with deterministic actions: in the saccadic network, the most probable of all 3 actions at every step; in the reaching network, the mode of each beta distribution at every step. A single representative model was selected for decoding experiments, however results extend to all models analyzed.

\subsection*{Evidence accumulation model}\label{EvAccMethods}
\addcontentsline{toc}{subsection}{Evidence accumulation model}

We evaluated agent performance against a lossless accumulation model with \textit{a priori} information about the dot motion stimulus to estimate the degree of temporal information used by the agent (see Fig \ref{fig:EvAccModel}).

First we constructed two 5 x 5 x 4 convolutional kernels for leftward and rightward motion at the ground truth of the dot motion stimulus (i.e., left or right motion with three interleaved frames at a speed of 1 pixel per frame). The stimulus was convolved with these kernels, output was summed across space (i.e., from local activation to global activation), and the difference between these leftward and rightward motion kernels was taken to approximate net motion energy \citep[see][]{adelson1985spatiotemporal, waskom2018perceptual}.

Approximated net motion energy was then accumulated for N steps of the motion stimulus for all coherence levels and directions. Accumulated values in each condition were normally distributed (KS tests; $p\text{s} > 0.05$). The mean accumulated value across all conditions was chosen as the signal detection theory threshold for left/right choices \citep{green1966signal}. Each condition for each N-step accumulation was simulated for 10,000 trials.

Results showed that the accuracy for this model was 50\%, 80.3\%, 86.2\%, and 89.1\% for 0, 1, 2, and 3 steps of accumulation, respectively. Trained 60 Hz saccadic reinforcement learning agents achieved an average of 93.30\% $\pm$ 0.37\% accuracy on the task across the same conditions—significantly higher than the hand-constructed accumulation model for 1 step of motion energy (80.3\%); $t(9) = 33.34, p = 4.84e\text{-}11$. These results suggest that the trained agents consider multiple samples of evidence over time in support of a decision, rather than considering single samples in isolation.

\subsection*{Decoding}
\addcontentsline{toc}{subsection}{Decoding}

Decoding analyses were inspired by studies decoding decisions in non-human primates based on neural activity \citep{kiani2014dynamics, peixoto2018population, peixoto2021decoding}. We trained a logistic regression classifier to predict the choices of a trained agent based on LSTM layer activity. We defined the decision variable (DV) as the log odds ratio of observing a particular choice ($T_{1}$: rightward, $T_{2}$: leftward) given the activity of all units considered ($r$):

\vspace{-10pt}
\begin{gather}
DV = log  \frac{P(T_{1}|r)}{P(T_{2}|r)} = \beta_{0} + \sum_{i=1}^{n}\beta_{i} r_{i}(t) \nonumber
\end{gather}

Where $r_{i}(t)$ are the z-scored activations for each unit, $\beta_{0}$ is an intercept term, and $\beta_{i}$ are the classifier weights. Classifier training finds a set of linear weights ($\beta_{i}$) on the activity of each unit that maximizes the probability of correctly predicting the agent’s choices. This can be viewed as finding the hyperplane that best separates neural activity according to observed choices. The distance of a point in high-dimensional activity space from this discriminant hyperplane can be viewed as the classifier’s degree of belief about the agent’s choice (i.e., the DV). Changes in the decoder’s prediction (i.e., a change of sign in the DV) before a response suggests a neural change of mind \citep{kiani2014dynamics, peixoto2018population, peixoto2021decoding}.

We trained a logistic regression classifier exclusively on the time step where agents made a response, and applied this trained classifier to all other time steps (both locked to dot motion stimulus onset, and response time). Trials with no responses were rejected, and data was randomly sampled so that an equal number of trials in each coherence condition remained. We performed 10-fold cross validation, stratified by response so that each fold had an equal number of left and right response trials for classification training. Chance decoding accuracy was defined as the response bias for the agent considered, which was slightly above 50\% for all agents (e.g., 50.9\% for the agent considered in Fig \ref{fig:fig4}).

Neural changes of mind are typically identified as the instance when a DV changes sign before a response on a single trial. In practice, DVs can be biased and noisy, requiring the addition of restrictions \citep{peixoto2021decoding}. Here we defined the change of mind threshold as the mean DV value at dot motion stimulus onset, instead of a change in sign (i.e., 2.2 instead of 0). We also required DVs to have exceeded this threshold by at least 10 DV units on the opposite side of space corresponding to the chosen side. In other words, if the agent ultimately chose the leftward option (corresponding to a negative DV), then a change of mind requires a DV that had at some point exceeded 12.2 (i.e., 10 + 2.2). Finally, we additionally required this threshold crossing to occur for at least 4 consecutive time steps to reduce changes of mind attributable to temporal noise. Decoded changes of mind results were qualitatively similar for the majority of these specific restriction values.

\subsection*{Human data}\label{humandata}
\addcontentsline{toc}{subsection}{Human data}

13 participants (7 women; Age: M = 19.29, SD = 1.44) took part in the experiment. All participants gave written consent prior to the experiment, which was approved by the University of Alberta’s Research Ethics Board. All participants were right-handed, had normal or corrected-to-normal vision, and did not know the purpose of the study. Participants were compensated with course credit.

Participants’ movements were recorded at 60 Hz using six Optitrak Flex 13 cameras (NaturalPoint, Inc., Corvallis, Oregon) mounted on two tripods, which tracked one passive, reflective motion-tracking marker placed near the tip of each participant’s right index finger. Stimuli were presented at 60 Hz (synchronized with the motion capture rate) on a vertical, table-mounted monitor (VIEWPixx/EEG; Saint-Bruno, Quebec). The position of the finger marker was co-registered in space with the monitor so the tabletop and monitor could be used as touch-interactive surfaces. Stimuli presentation and data collection were controlled with MATLAB using Psychtoolbox \citep{kleiner2007s}.

Participants were seated in a semi-dark room with a computer monitor at a viewing distance of 57 cm. Participants were instructed to maintain their gaze on the central red fixation square during every trial. Stimuli were presented on a black background (see Fig \ref{fig:HumanExperiment}).

On each trial, participants saw a random dot motion stimulus and were asked to report the net motion of the left/right moving dots as quickly and accurately as possible. White dots (3 x 3 pixel squares) were presented within a 5° circular aperture at the center of the screen. Dots moved at a speed of 5 deg/s, and average dot density was set at 16.8 dots/degree\textsuperscript{2}/s. As in \citet{kiani2013integration}, the stimulus consisted of three independent and interleaved sets of dots presented on successive video frames. Motion direction (left, right) and coherence (0\%, 3.2\%, 6.4\%, 12.8\%, 25.6\%, 51.2\%, and 100\%) varied pseudorandomly from trial-to-trial. Dot motion within each trial continued from stimulus onset until one of the two targets had been selected.

Throughout the trial, two gray circles (3 degrees\textsuperscript{2}) to the left and right of center fixation (10 degrees) were visible at a height of ~30 cm relative to the tabletop. Participants responded by reaching to touch one of the two circles with their right index finger. Reaching distance from start position to targets was approximately 30 cm forward, and 10 cm laterally. No feedback was given on dot motion trials so post-decision confidence could be rated.

At the end of each trial, participants returned their finger to the start position to rate their confidence in their decision. A white horizontal line (30 cm long) with the text, ``How confident?” appeared on the screen, and participants were asked to reach and touch a point on the line corresponding to their confidence in the previous decision (left, 0\% confident; right, 100\% confident). Once participants had finished adjusting the confidence slider, they could return to the start position to begin the next trial.

Intertrial intervals were randomly drawn from a truncated exponential distribution (range: 700-1000 ms; mean: 820 ms; as in \citealp{resulaj2009changes}). Participants completed 10 practice trials, followed by 10 blocks of 39 trials for a total of 400 trials per participant. Trials were excluded for reaction times less than 150 ms or greater than 6 s, movement times greater than 2 s, confidence rating times greater than 5 s, or for motion tracking recording errors (\citealp{gallivan2014three}; trial rejection: M = 15.61\%, SD = 7.64\%).

\newpage
\section*{Supplementary material}\label{sup} 
\addcontentsline{toc}{section}{Supplementary material}

\setcounter{figure}{0}
\renewcommand{\figurename}{Figure}
\renewcommand{\thefigure}{S\arabic{figure}}
\setcounter{table}{0}
\renewcommand{\thetable}{S\arabic{table}}

\subsection*{Ablations} \label{supp_ablations}
\addcontentsline{toc}{subsection}{Ablations}

A set of agents (N = 9) were trained in line with the 60 Hz saccadic agents described in Fig \ref{fig:fig2}. Here the only difference was in the training environment—agents were trained only on noiseless dots (i.e., only coherences of 100\%). Theory predicts that noisy environments are necessary for accumulation mechanisms to emerge. On average, we find this to be the case; average agent performance at the end of training was not significantly different from the one-sample accumulation threshold (80.3\%) identified by the evidence accumulation model, $t(8) = 0.10, p = 0.54$.

A second set of agents (N = 9) were trained in line with the 60 Hz saccadic agents described in Fig \ref{fig:fig2}. Here the only difference was in agent architecture—agents had a fully-connected layer in place of a recurrent LSTM layer with the same number of units. Note that because the recurrent agents also have parameters associated with their recurrent connections, the recurrent agents have more parameters than non-recurrent agents with a fully-connected layer in place of a LSTM layer. Average non-recurrent agent performance at the end of training was significantly higher than the one-sample accumulation threshold identified by the evidence accumulation model, $t(8) = 41.11, p = 6.75e\text{-}11$. In addition, non-recurrent agents learned the random dot motion task faster than recurrent agents, and displayed similar choice and response time sensitivity as their recurrent counterparts. While these agents did display dynamics that reflected momentary motion evidence in their CNN units, LSTM layer activation patterns strongly supported a non-primate-like extrema detection mechanism instead of a primate-like evidence accumulation mechanism. Recall that using an extrema detection mechanism, a decision maker waits until an individual sample is large enough to trigger a response \citep{stine2020differentiating}. Under this mechanism, agents base their decisions on a single time step of evidence, rather than the accumulated history of evidence in time.

Overall, these two ablations support the idea that environmental noise and recurrent connections are necessary ingredients for primate-like decision making to reliably emerge via reinforcement learning.

\begin{figure}[!htb]
\centering
\includegraphics[width=1.0\textwidth]{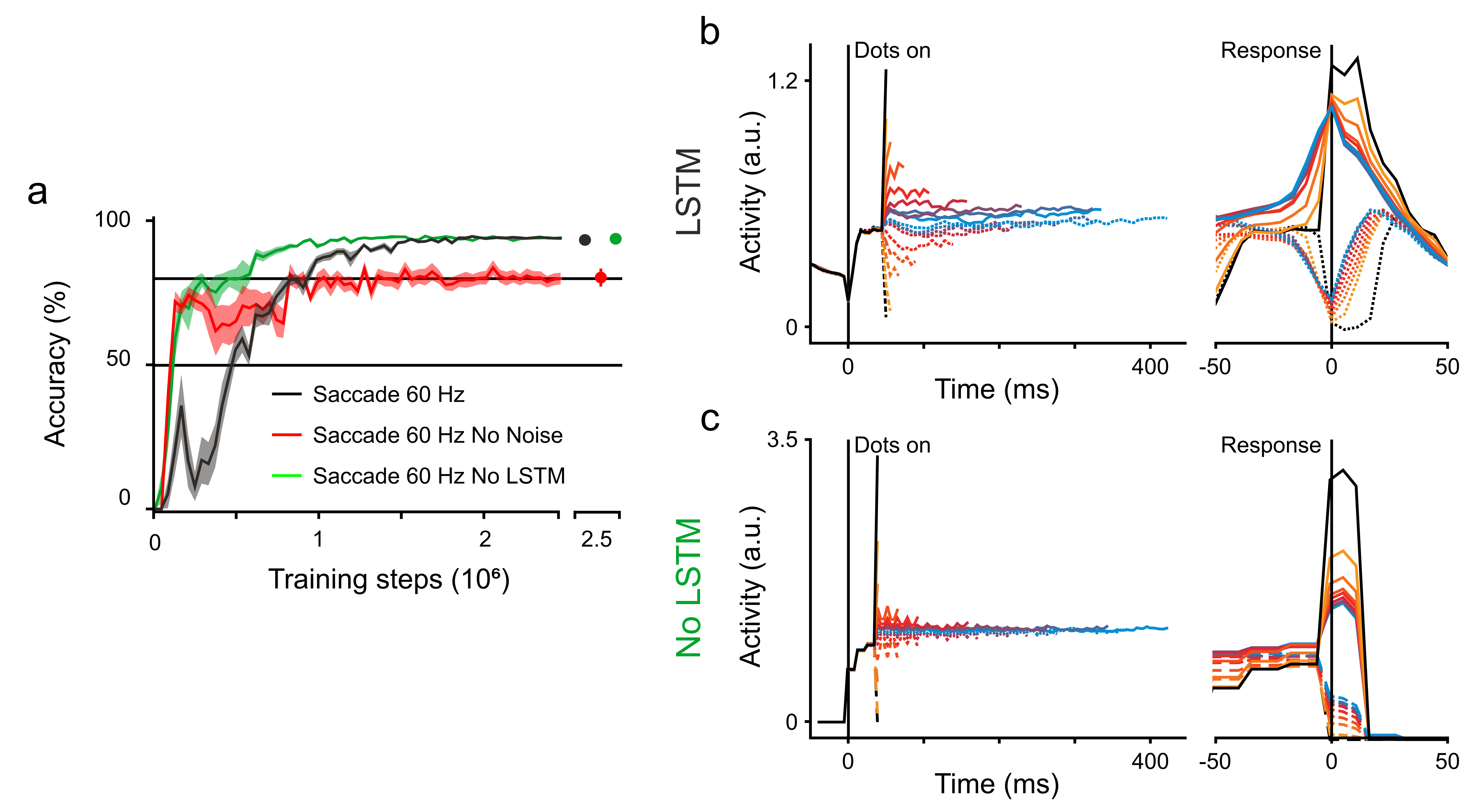}
\caption{
    Ablation results. Agents trained without environmental noise do not on average exceed the one-sample accuracy threshold determined from the evidence accumulation model. Agents trained without recurrence (i.e., no LSTM) display primate-like behaviour, but \textit{not} primate-like mechanisms.
    a) Accuracy during and after training. Agents trained without environmental noise (red), do not at any point exceed the one-sample 80.3\% accuracy threshold. In contrast, agents trained without recurrence (green) exceed this threshold even earlier than recurrent agents (black).
    b) Single-unit LSTM dynamics from a representative 180 Hz agent (black in a) aligned to stimulus onset (left) and response (right). Reproduced from Fig \ref{fig:fig3}c. Note the gradual accumulation of activity proportional to coherence before a response (right).
    c) Single-unit dynamics from a representative 180 Hz non-recurrent agent (green in a; ``No LSTM") aligned to stimulus onset (left) and response (right). Unit is from the fully-connected layer which replaced the LSTM layer for these agents. Since behaviour (the first key property of primate-like decision making) was similar between agents, we trained 180 Hz agents with and without recurrence to interrogate dynamics (the second key property of primate-like decision making). Note the response-aligned dynamics for this non-recurrent agent suggest that decision information does not accumulate gradually before a response (as in b), but only on the single time step before a response, suggesting an alternative extrema detection mechanism.
}
\label{fig:ablations}
\end{figure}

\newpage

\subsection*{Unit selectivity} \label{sup_units} 
\addcontentsline{toc}{subsection}{Unit selectivity}

We determined unit selectivity for CNN units by performing a linear regression of mean unit activity against coherence for each motion direction independently (i.e., one regression for leftward motion, and one for rightward motion). Significant linear regressions at a level of $p < 0.05$ were interpreted as units that were selective for that direction of motion. LSTM selectivity was performed instead on the difference between activity at stimulus onset and activity after the first full step of dot motion. Analysis was performed for all 10 trained agents in the 180 Hz saccade task.

Several CNN units were significantly selective for only leftward ($4.2\% \pm 0.7\%$), only rightward ($7.5\% \pm 1.1\%$), or both directions of motion ($63.3\% \pm 2.1\%$), while some units were not selective for either direction ($25.0\% \pm 1.4\%$). Several LSTM units were significantly selective for only leftward ($3.6\% \pm 0.5\%$), only rightward ($3.3\% \pm 0.5\%$), or both directions of motion ($66.0\% \pm 1.2\%$), while some units were not selective for either direction ($27.1\% \pm 1.3\%$).

\begin{figure}[!htb]
\centering
\includegraphics[width=1.0\textwidth]{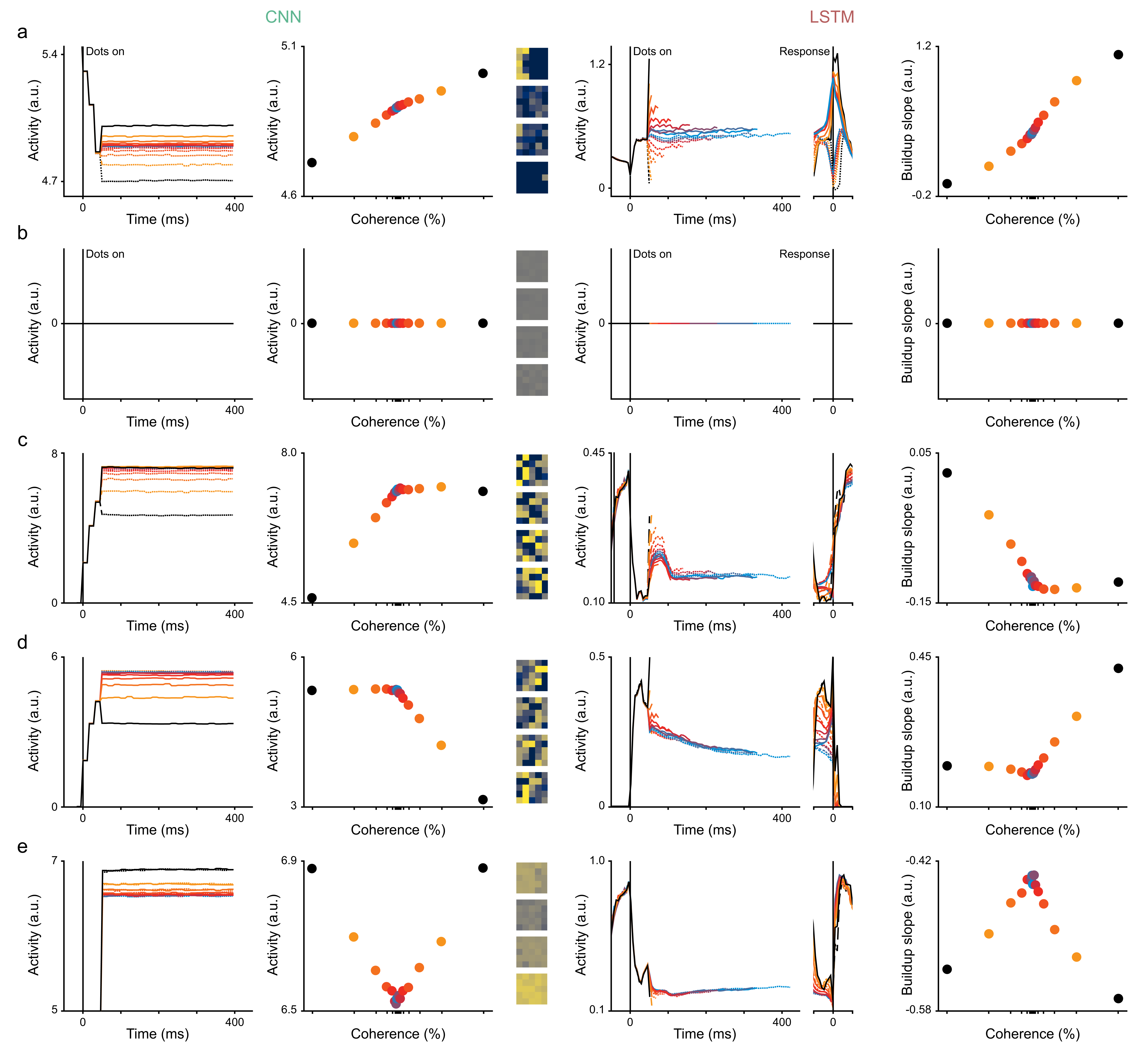}
\caption{
Example CNN and LSTM unit selectivity profiles from a representative 180 Hz saccade agent. From left to right, the CNN unit activity over time, the CNN unit activity at a single point in time used for unit selectivity analyses, the 5 x 5 x 4 CNN kernel for this unit, the LSTM unit activity over time relative to stimulus onset, the LSTM unit activity over time relative to response, and the LSTM unit activity slope used for unit selectivity analyses.
a) Units shown in Fig 3.
b) Units with no selectivity.
c) Units selective only for leftward motion.
d) Units selective only for rightward motion.
e) Units selective for both motion directions in the same way.
}
\label{fig:unitselectivity}
\end{figure}

\begin{figure}[!htb]
\centering
\includegraphics[width=1.0\textwidth]{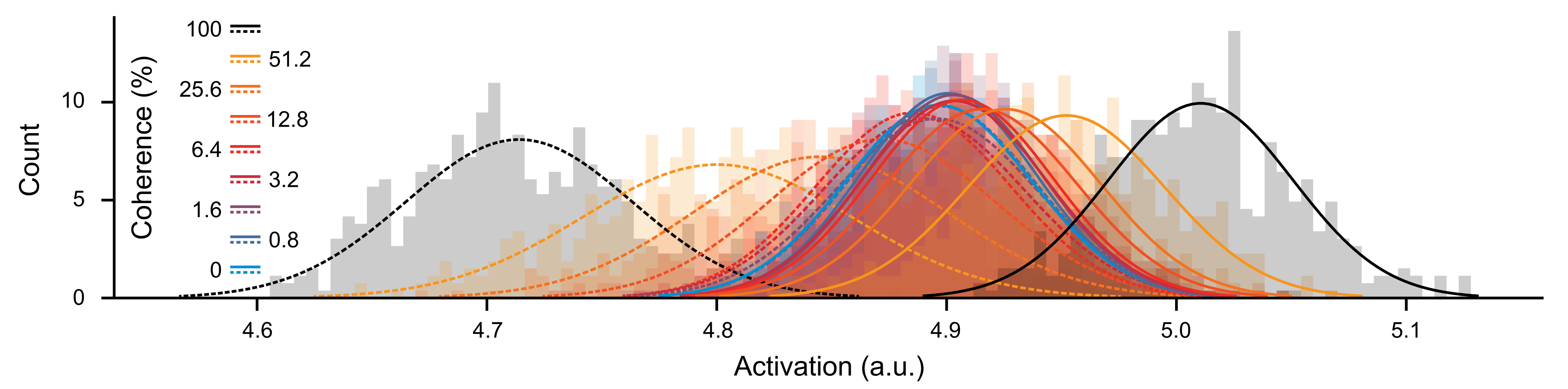}
\caption{
Distribution of activity for the example CNN unit in Fig \ref{fig:fig3} from a single time step on independent evaluation trials. Gaussian distributions were fit to activity within each direction-coherence condition. Distributions of CNN unit activity were normally distributed within each direction-coherence condition (KS tests; $p\text{s} > 0.05$), in line with the hand-crafted evidence accumulation model (see \nameref{methods}), standard evidence accumulation models \citep{gold2007neural}, and recordings from primate area MT \citep{britten1992analysis}.
}
\label{fig:CNNdistribution}
\end{figure}

\begin{figure}[!htb]
\centering
\includegraphics[width=1.0\textwidth]{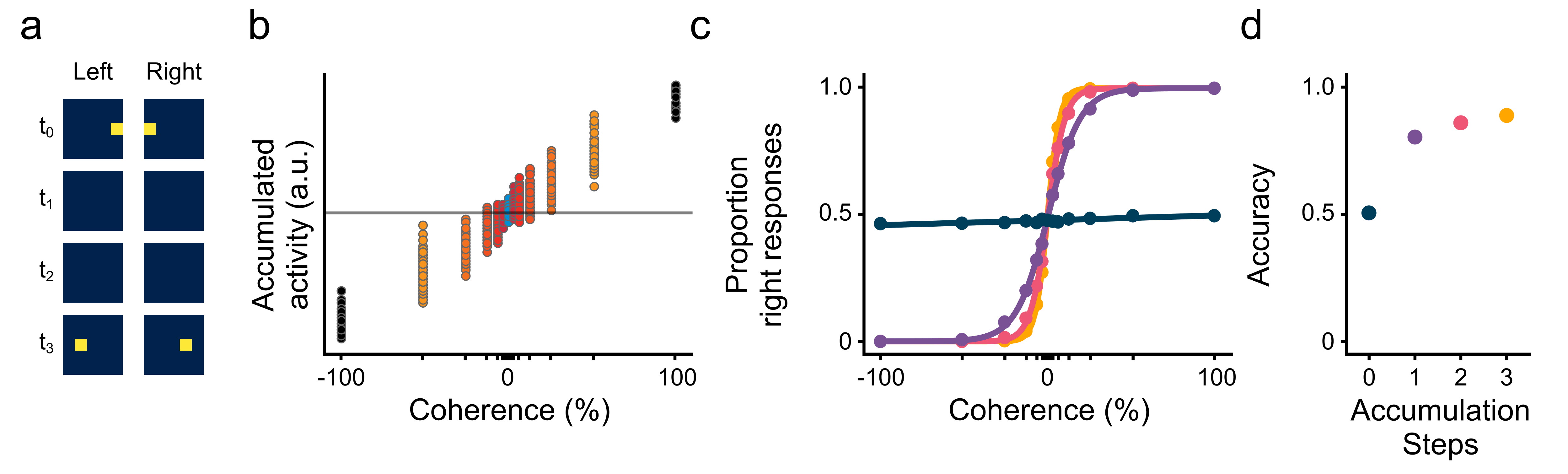}
\caption{
Evidence accumulation model and results.
a) Two 5 x 5 x 4 convolutional kernels were hand-constructed with \textit{a priori} knowledge of the dot motion stimulus—one for leftward motion and one for rightward motion at a speed of 3 pixels with 3 interleaved frames. The dot motion stimulus was convolved with each of these two kernels and summed across space. The difference between these two values was taken as a proxy for net motion energy.
b) This proxy for net motion energy was accumulated for n steps, and a left/right decision was made based on a signal detection threshold. Each dot represents the value for 3 steps of accumulated net motion energy on an independent simulated trial.
c) Model accuracy for 0, 1, 2, and 3 steps of accumulation. Accuracy increases with more steps of accumulation. See d for color legend.
d) Accuracy averaged across coherence conditions. The hand-constructed evidence accumulation model achieved 50\% accuracy for 0 steps of accumulation, 80.3\% accuracy for 1 step of accumulation, 86.2\% accuracy for 2 steps of accumulation, and 89.1\% accuracy for 3 steps of accumulation.
}
\label{fig:EvAccModel}
\end{figure}

\begin{table}[!h] 
  \centering
  \begin{tabular}{lll}
    \toprule
    &\multicolumn{2}{c}{Task/Network}                   \\
    \cmidrule(r){2-3}
    Hyperparameter     & Saccade     & Arm \\
    \midrule
    No. of Conv kernels & 64  & 64     \\
    No. of LSTM units     & 128 & 128      \\
    No. of environments & 16 & 16 \\
    Steps per rollout & 256 & 256 \\
    No. minibatches & 2 & 8 \\
    Simulation rate (Hz) & 60 \& 180 & 60  \\
    Steps before motion onset & (2, 11] & (0, 1]  \\
    Max time per trial (s) & 3 \& 2 & 3  \\
    Adam learning rate & 3e-4 $\rightarrow$ 0.0 & 3e-4 $\rightarrow$ 0.0  \\
    Discount rate ($\gamma$) & 0.99 & 0.99  \\
    PPO clip ($\epsilon$) & 0.2 $\rightarrow$ 0.0 & 0.2 $\rightarrow$ 0.0  \\
    PPO critic coef. ($\beta_{V}$) & 0.25 & 0.25  \\
    PPO entropy coef. ($\beta_{ent}$) & 0 & 0  \\
    GAE parameter ($\lambda$) & 0.95 & 0.95 \\
    Global norm grad clip & 0.5 & 0.5  \\
    Move forward reward coef. & N/A & 0.005  \\
    \bottomrule
  \end{tabular}
\caption{Hyperparameters for artificial agents responding via simulated saccades or reaches. Proximal policy optimization (PPO) algorithm hyperparameters for recurrent agents are described in \citet{schulman2017proximal}.}
\label{hyperparam-table}
\end{table}

\begin{figure}[!htb]
\centering
\includegraphics[width=1.0\textwidth]{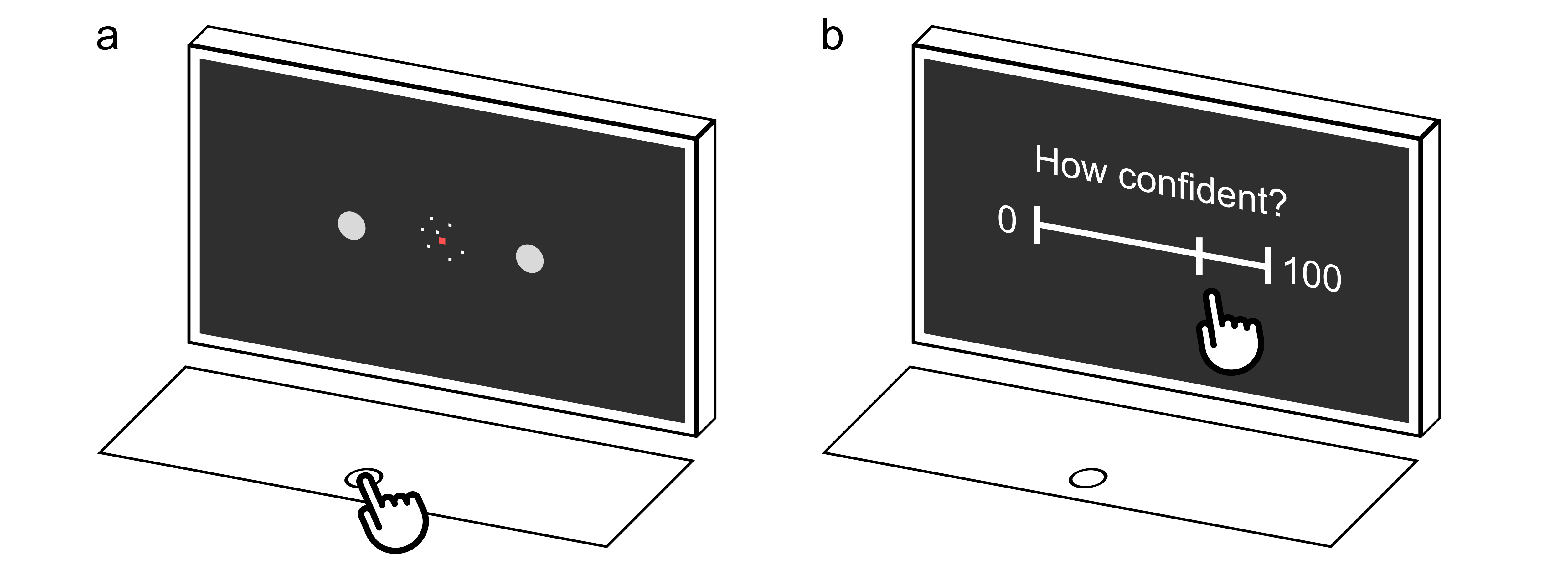}
\caption{
Human experiment timeline.
a) Human participants were shown the random dot motion task on an upright screen while seated at a desk. Participants were asked to fixate on a red square in the middle of the monitor, and responded by moving their finger from the start position in front of them to one of the two grey targets on the monitor. Feedback about whether their decision was correct was withheld from the participants.
b) After each trial, participants were asked to rate how confident they were in their decision on a sliding scale from 0 to 100 before continuing to the next trial.
}
\label{fig:HumanExperiment}
\end{figure}

\begin{figure}[!htb]
\centering
\includegraphics[width=0.6\textwidth]{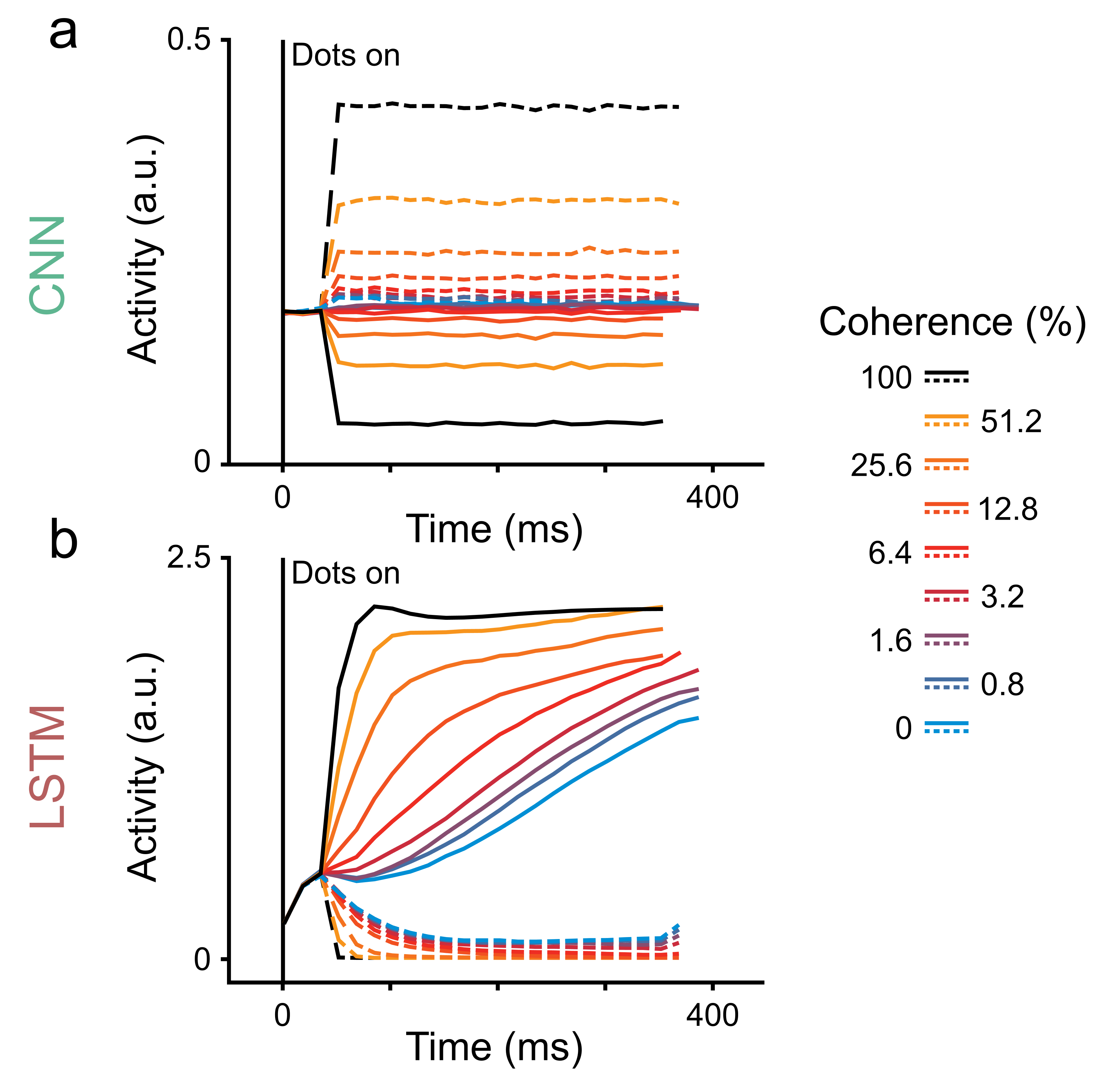}
\caption{
Internal dynamics from a representative 60 Hz reaching agent. Solid lines indicate rightward motion conditions, and dashed lines indicate leftward motion conditions. For 0\% coherence conditions (blue), traces are separated by when the agent ultimately chose the left or right target. Traces are averages of each condition for the agent.
a) Dynamics from an example CNN unit.
b) Dynamics from an example LSTM unit.
}
\label{fig:armdynamics}
\end{figure}

\end{document}